\documentclass[aps,prc,reprint,amsmath,nofootinbib,superscriptaddress]{revtex4-1}

\usepackage{hyperref}
\usepackage{graphicx}
\usepackage{amsmath}
\usepackage{mathtools}
\usepackage{paralist}

\usepackage{xcolor}
\usepackage[normalem]{ulem}
\makeatletter
\def\uwave{\bgroup \markoverwith{\lower3.5\p@\hbox{\sixly \textcolor{red}{\char58}}}\ULon}
\font\sixly=lasy6 
\makeatother
\usepackage{mdwlist}

\begin{document}

\title{Hydrodynamic simulations of relativistic heavy-ion collisions\\ with different lattice QCD calculations of the equation of state}

\author{J.\ Scott Moreland}
\affiliation{Department of Physics, Duke University, Durham, NC 27708-0305}
\author{Ron A.\ Soltz}
\affiliation{Lawrence Livermore National Laboratory, Livermore, CA 94551-0808}

\date{\today}

\begin{abstract}
Hydrodynamic calculations of ultra-relativistic heavy ion collisions are performed using the iEBE-VISHNU 2+1D code with fluctuating initial conditions and three different parameterizations of the Lattice QCD equations of state: continuum extrapolations for stout and HISQ/tree actions, as well as the s95p-v1 parameterization based upon calculations using the p4 action.  All parameterizations are matched to a hadron resonance gas equation of state at $T=155$~MeV, at which point the calculations are continued using the UrQMD hadronic cascade. Simulations of $\sqrt{s_{NN}}=200$ GeV Au+Au collisions in three centrality classes are used to quantify anisotropic flow developed in the hydrodynamic phase of the collision as well as particle spectra and pion HBT radii after hadronic rescattering which are compared with experimental data. Experimental observables for the stout and HISQ/tree equations of state are observed to differ by less than a few percent for all observables, while the s95p-v1 equation of state generates spectra and flow coefficients which differ by $\sim$10--20\%. Calculations in which the HISQ/tree equation of state is sampled from the published error distribution are also observed to differ by less than a few percent.
\end{abstract}

\maketitle

\section{Introduction}

Quantum Chromodynamics (QCD) predicts that at sufficiently high temperature or density, nuclear matter exists in a deconfined state of quarks and gluons known as a quark-gluon plasma (QGP). 
This state of matter filled the early universe several microseconds after the big bang and is now recreated and studied in the laboratory by colliding heavy ions at relativistic energies at the Relativistic Heavy Ion Collider (RHIC) and the Large Hadron Collider (LHC).

Quantitative model to data comparison, using simulations based on relativistic hydrodynamics, is the optimal means to extract properties of QGP produced by relativistic heavy-ion collisions which expands and freezes into hadrons too quickly for direct observation. 
These hydrodynamic descriptions require two essential ingredients to specify the full time evolution of the QGP fireball: initial conditions which describe the thermal profile of the QGP droplet at some early starting time and a QCD Equation of State (EoS) which interrelates the energy density, pressure and temperature of each fluid cell in local thermal equilibrium.

Lattice discretization is the only reliable method to calculate the QCD equation of state at zero baryochemical potential in the vicinity of the QGP phase transition and hence constitutes a critical component of hydrodynamic simulations. 
While lattice techniques are rigorous in their treatment of the underlying QCD Lagrangian, they are subject to statistical and systematic errors inherent in the lattice discretization procedure. 
These errors are manifest in differences in the continuum extrapolated QCD trace anomaly and lead to an overall uncertainty in the true value of the QCD equation of state.

To date there have been few sensitivity studies on the influence of the EoS on hydrodynamic simulation results. 
These have been limited to studies of the order of the phase transition~\cite{Huovinen:2005gy}, different parameterization schemes for the LQCD EoS \cite{Huovinen:2009yb} and data driven Bayesian techniques to constrain parameterizations of the EoS motivated by LQCD calculations \cite{Pratt:2015zsa,Sangaline:2015isa}.
However, a sensitivity study on the inherent errors in the LQCD EoS has not yet been performed, primarily because continuum extrapolations for the LQCD EoS at zero baryon density have only recently become available \cite{Borsanyi:2013bia,Bazavov:2014pvz}. 
In this work, we quantify the effect of lattice errors on simulations of relativistic heavy-ion collisions by comparing simulation predictions obtained with QCD EoS calculations by the Wuppertal-Budapest collaboration using the stout fermion action \cite{Borsanyi:2013bia} and the HotQCD collaboration using the HISQ/tree action \cite{Bazavov:2014pvz}.  
We also compare to the older s95p-v1 parameterization \cite{Huovinen:2009yb} constructed from calculations performed on coarser ($32^3 \times 8$) lattices using p4 and asqtad actions without continuum extrapolation \cite{Bazavov:2009zn}.  
The equations of state are analyzed using a modern event-by-event hybrid simulation which couples viscous hydrodynamics to a hadronic afterburner to calculate flows, spectra and Bertsch-Pratt radii and are compared to measurements at the Relativistic Heavy-Ion Collider (RHIC).
We also perform a set of calculations in which the HISQ/tree continuum EoS is sampled from within the published error range.

\section{Equations of State}
\label{eos}

\begin{figure}[t]
  \includegraphics[width=\columnwidth]{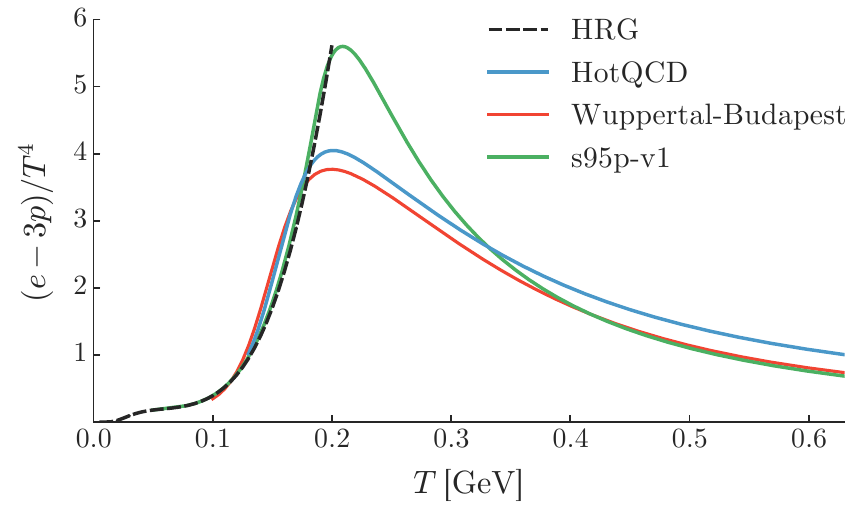}
  \caption{\label{fig:trace} The QCD interaction measure for a hadron resonance gas (HRG) alongside recent lattice calculations from the HotQCD and Wuppertal-Budapest collaborations as well as the older s95p-v1 lattice parameterization \cite{Bazavov:2014pvz, Borsanyi:2013bia, Huovinen:2009yb}.}
\end{figure}

The Wuppertal-Budapest, HotQCD and s95p-v1 EoS parameterizations used in this work all employ staggered fermion actions with varying level improvements: additional terms added to remove lattice artifacts and improve simulation convergence. 
For example, both the stout and HISQ/tree actions used by the Wuppertal-Budapest and HotQCD calculations contain additional smearing of the gluon links relative to the p4 action used to construct the s95p-v1 parameterization.
Moreover, the Wuppertal-Budapest stout action omits second order corrections in the lattice spacing which are common to the other three. 

The three analyses are further distinguished by the granularity of the lattices used in each calculation.
The p4 results used in the s95p-v1 parameterization are from ($32^3 \times 8$) lattices, referred to by the number of temporal dimension, $N_{\tau}=8$, while the HISQ/tree continuum extrapolation was calculated for $N_{\tau}=8$, 10, and 12, and the stout results for lattices with $N_{\tau}=6$, 8, 10 and 12. 
For a more detailed discussion of the EoS calculations and relative improvements of the staggered fermion actions see \cite{Soltz:2015ula}.

LQCD EoS calculations are obtained from the trace of the stress-energy tensor, equal to the difference between the energy density and three times the pressure. 
This quantity is typically referred to as the interaction measure or trace anomaly  because it measures deviations from the conformal equation of state. 
Scaled by the fourth power of the temperature, the trace anomaly forms a dimensionless measure 
\begin{equation}
 I \equiv \frac{\Theta^{\mu\mu}(T)}{T^4} = \frac{e - 3p}{T^4},
\end{equation}
where $\Theta$ is the stress-energy tensor, $e$ is the local fluid energy density, $p$ the pressure and $T$ the temperature.

Lattice calculations typically extend down to temperatures of ${\sim}130$~MeV, where small deviations with the Hadron Resonance Gas (HRG) EoS may begin to develop.
This is evident in Fig.~\ref{fig:trace}, which shows the trace anomaly of the HRG EoS alongside results from the HotQCD and Wuppertal-Budapest collaborations with HISQ/tree and stout actions respectively, as well as the older s95p-v1 parameterization obtained using the p4 action.  
Both the HISQ/tree and stout EoS results begin to pull away from the HRG EoS at temperatures above $130$~MeV, while the s95p-v1 parameterization agrees with the HRG results up to a matching temperature of $183.8$~MeV by construction.

\begin{figure}[t]
  \includegraphics[width=\columnwidth]{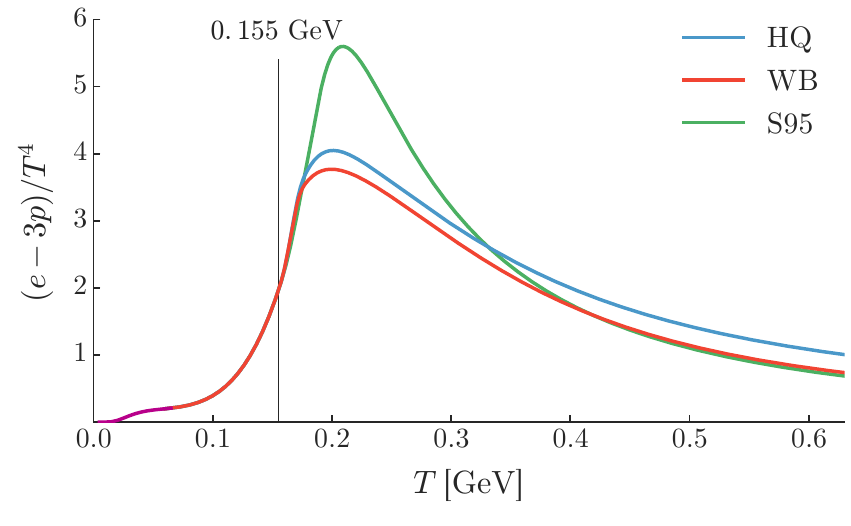}
  \caption{\label{fig:trace_final} The modified QCD interaction measures for the HQ, WB and S95 EoS obtained from equation \eqref{interaction} and the corresponding lattice
	  parameterizations in Fig.~\ref{fig:trace}. The vertical line marks the hydro-to-micro switching temperature $T_\text{sw} = 155$~MeV.}
\end{figure}

Although both the Wuppertal-Budapest and HotQCD collaborations have provided parameterizations suitable for insertion into hydrodynamic codes, the matching temperature of 130~MeV falls below the 155--165~MeV temperature range where hybrid simulations typically switch from relativistic viscous hydrodynamics to a microscopic kinetic description such as UrQMD~\cite{Bass:1998ca,Bleicher:1999xi}.  
We note also that recent estimates for the freeze-out temperature derived from combining lattice calculations and experimental data also fall within this range~\cite{Bazavov:2014xya, Adare:2015aqk}.  
To ensure a self consistent description of the collision dynamics where the simulation switches from hydrodynamics to microscopic transport, we modify each lattice EoS to match the HRG EoS at the desired hydro-to-micro switching temperature. We thus define a new piecewise interaction measure
\begin{equation}
 \label{interaction}
 I(T) =
  \begin{cases}
   I_\text{hrg}(T)	& T \le T_1, \\
   I_\text{blend}(T)	& T_1 < T < T_2, \\ 
   I_\text{lattice}(T)	& T \ge T_2,
  \end{cases}
\end{equation}
where $I_\text{hrg}$ and $I_\text{lattice}$ are the HRG and LQCD trace anomalies pictured in Fig.~\ref{fig:trace}, and $I_\text{blend}$ is a function 
\begin{equation}
  \label{interpolation}
  I_\text{blend} = (1-z)\, I_\text{hrg} + z\, I_\text{lattice}
\end{equation}
which smoothly connects between the two in the temperature interval $T_1 < T < T_2$. The interpolation parameter $z \in [0,1]$ is constructed to match the first and second derivatives at the endpoints of the interpolation interval,
\begin{align}
 \label{smoothstep}
  z &= 6 x^5 - 15 x^4 + 10 x^3 \\
  \text{where } x &= (T - T_1)/(T_2 - T_1).
\end{align}
We fix the boundaries of the blending region $T_1=155$~MeV and $T_2=180$~MeV to impose matching at the switching temperature $T_\text{sw} = 155$~MeV, which coincides with the pseudo-critical phase transition temperatures of the HotQCD and Wuppertal-Budapest EoS~\cite{Borsanyi:2010gc,Bazavov:2012iu,Bhattacharya:2014iw}. The modified interaction measures, hereafter referred to simply as HQ, WB and S95, are plotted in Fig.~\ref{fig:trace_final}\,. This interpolation procedure imposes the necessary matching condition on either side of the switching temperature (vertical line) with minimal disturbance to the peak of the LQCD trace anomaly at higher temperatures.

\begin{figure}[b]
  \includegraphics[width=\columnwidth]{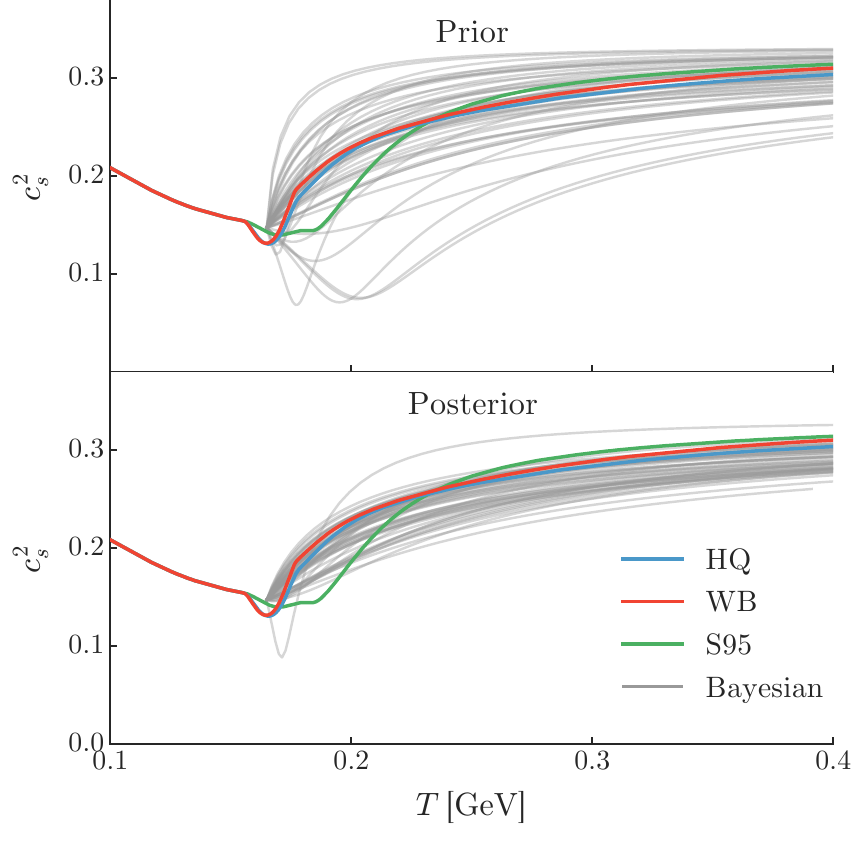}
  \caption{\label{fig:cs} Squared speed of sound $c_s^2$ plotted versus temperature $T$ for the HQ, WB and S95 equations of state pictured in 
           Fig.~\ref{fig:trace_final}. Top panel shows EoS parameterizations from the Bayesian prior used in reference \cite{Pratt:2015zsa} (thin grey lines)
           while the bottom panel shows samples from the Bayesian posterior once the samples have been constrained by experimental data.
          }
\end{figure}

Signal propagation in the QGP medium is characterized by the speed of sound, expressed in terms of the pressure and energy density as $c_s^2 = dp/de$.
In Fig.~\ref{fig:cs} we plot the squared speed of sound for the HQ, WB and S95 interaction measures shown in Fig.~\ref{fig:trace_final} alongside recent results from a systematic Bayesian analysis used to constrain parameterized forms of the LQCD EoS by simultaneously fitting model predictions to multiple observables at RHIC and the LHC \cite{Pratt:2015zsa}. 
The top panel of Fig.~\ref{fig:cs} shows the three lattice parameterizations used in this work plotted against 50 parametric EoS samples (thin grey lines) from the Bayesian prior, while the bottom panel of Fig.~\ref{fig:cs} shows the same lattice results plotted against samples from the Bayesian posterior, i.e.\ once the EoS curves have been constrained by data. The more tightly clustered posterior curves show a clear preference for the present lattice results.  Although these constraints are not able to resolve differences between the different lattice calculations, they fall below the continuum extrapolations for temperatures above 0.2~GeV.

Within the three lattice calculations used in this study, the HQ and WB speed of sound curves are in good agreement while the S95 parameterization remains softer in a wider interval about the QGP phase transition. We note that the parametric transition \eqref{interpolation} modifies the speed of sound in the vicinity of the EoS matching temperature but is constructed to preserve continuity across the desired transition region.

With the trace anomalies in hand, the energy density, pressure and entropy density are easily interrelated to specify the equation of state used in the analysis,
\begin{align}
 \label{conversion}
 \frac{p(T)}{T^4} &= \int_0^T dT'\, \frac{I(T')}{T'}, \\
 \frac{e(T)}{T^4} &= I(T) + 3\, \frac{p(T)}{T^4}, \\
 \frac{s(T)}{T^3} &= \frac{e(T) + p(T)}{T^4}. 
\end{align}
 
For clarity, Figs.~\ref{fig:trace}--\ref{fig:cs} do not include the respective errors bands for the HotQCD and Wuppertal-Budapest trace anomalies, but both calculations devote considerable effort to providing an accurate error estimate for their respective calculations~\cite{Borsanyi:2013bia,Bazavov:2014pvz}.  Common contributions to the errors
come from variations in spline fits to the interaction measures, differences between quadratic and quartic extrapolations in the lattice spacing, and
small (~2\%) variations in the temperature scale.  Errors are typical of order 5\% for most quantities, and increase to 5--10\% in the transition region
where the curves are steepest.

\section{Hybrid Model}

The equations of state are embedded in the event-by-event iEBE-VISHNU hybrid model which uses the VISH2+1 boost-invariant viscous hydrodynamics code \cite{Song:2007ux} to simulate the 
time evolution of the QGP medium and the microscopic UrQMD hadronic afterburner \cite{Bass:1998ca, Bleicher:1999xi} for subsequent evolution below the QGP transition temperature. Where necessary, free parameters of the model are tuned to facillitate model-to-data comparison with $\sqrt{s_{NN}}=200$ GeV gold-gold collisions at RHIC. In this section, we briefly outline
the implementation of the model used in the analysis; for a more detailed explanation of the model see reference \cite{Shen:2014vra}.

\subsection{Initial conditions}
\label{initial_condition}

The initial conditions represent the largest source of uncertainty in current hydrodynamic simulations and a number of models exist in the literature which have described
the experimental data with varying degrees of success \cite{Schenke:2012wb, Niemi:2015qia, Chatterjee:2015aja, Moreland:2014oya, Drescher:2006pi, Adler:2013aqf}. Because the goal of the present work is to measure the sensitivity of the hydrodynamic evolution to differences in the QGP EoS and \emph{not} to obtain the overall best fit of model to data, we choose the simplest and most widely adopted initial condition implementation based on a two-component Glauber model; for an overview see \cite{Miller:2007ri}.

In the two-component ansatz, initial entropy is deposited proportional to a linear combination of nucleon participants and binary nucleon-nucleon collisions,
\begin{equation}
 dS/dy \,\vert_{y=0} \propto \frac{(1-\alpha)}{2}N_\text{part} + \alpha N_\text{coll}
 \label{twocomponent}
\end{equation}
where for the binary collision fraction, we use $\alpha=0.14$ which has been shown to provide a good description of the centrality dependence of charged particle 
multiplicity in $\sqrt{s_{NN}}=200$ GeV gold-gold collisions \cite{Shen:2014sfi}.

The entropy is localized about each nucleon's transverse parton density $T_p({\bf x})$,
\begin{align}
 dS/dy \,\vert_{y=0} &\propto \smashoperator{\sum_{i=0}^{N_\text{part,A}}} w_i\, T_p({\bf x} - {\bf x}_i)(1-\alpha + \alpha\, N_\text{coll,i}) \nonumber \\
                     &+ \smashoperator{\sum_{j=0}^{N_\text{part,B}}} w_j\, T_p({\bf x} - {\bf x}_i)(1-\alpha + \alpha\, N_\text{coll,j})
 \label{glauber}
\end{align}
where the summations run over the participants in each nucleus, $N_\text{coll,i}$ denotes the number of binary collisions suffered by the $i$-th nucleon 
and the proton density $T_p({\bf x})$ is described by a Gaussian
\begin{equation}
 T_p({\bf x}) = \frac{1}{\sqrt{2 \pi B}} \exp \left(-\frac{x^2+y^2}{2 B} \right)
\end{equation}
with transverse area $B = 0.36$ $\text{fm}^2$.

The random nucleon weights $w_i$ in equation \eqref{glauber} are sampled independently from a Gamma distribution with unit mean
\begin{equation}
 P_k(w) = \frac{k^k}{\Gamma(k)} w^{k-1} e^{-k w},
\end{equation}
and shape parameter $k = \text{Var}(P)^{-1}$ which modulates the variance of the distribution. 
Such fluctuations are typically added to reproduce the large multiplicity fluctuations observed in minimum bias proton-proton collisions \cite{Adare:2008ns, Dumitru:2012yr, Moreland:2012qw, Bozek:2013uha, Shen:2014sfi}. 
In this work the shape parameter is fixed to $k=1$ determined by a fit to the $\sqrt{s}=200$ GeV UA5 proton-antiproton data \cite{Ansorge:1988kn}. 

The initial condition profiles, which provide the entropy density at the QGP thermalization time, are finally rescaled by an overall normalization factor to fit the measured charged particle multiplicity in 0--10\% centrality collisions.

\subsection{Hydrodynamics and Boltzmann transport}

The hydrodynamic equations of motion are obtained in the iEBE-VISHNU model by solving the second-order Israel-Stewart equations,
\begin{equation}
 \partial_\mu T^{\mu\nu} = 0, \quad T^{\mu\nu} = e u^\mu u^\nu - (p + \Pi) \Delta^{\mu\nu} + \pi^{\mu\nu},
\end{equation}
where the bulk pressure $\Pi$ and shear stress $\pi^{\mu\nu}$ satisfy the relaxation equations,
\begin{align}
 \label{viscosity}
  \mathcal{D}\Pi = &- \frac{1}{\tau_\Pi}(\Pi + \zeta \theta) - \frac{1}{2} \Pi \frac{\zeta T}{\tau_\Pi}d_\lambda \left(\frac{\tau_\Pi}{\zeta T} u^\lambda \right), \nonumber \\
  \Delta^{\mu\alpha} \Delta^{\nu\beta} \mathcal{D}\pi_{\alpha\beta} = &- \frac{1}{\tau_\pi}(\pi^{\mu\nu} - 2 \eta \sigma^{\mu\nu}) \\
  &- \frac{1}{2} \pi^{\mu\nu} \frac{\eta T}{\tau_\pi} d_\lambda\left(\frac{\tau_\pi}{\eta T} u^\lambda \right ).
\end{align}

We follow the work in reference \cite{Shen:2014sfi} and fix the bulk viscosity $\zeta$ and shear viscosity $\eta$ in equation \eqref{viscosity} using a constant specific shear viscosity $\eta/s=0.08$ 
and vanishing bulk viscosity $\zeta/s=0$ in the hydrodynamic phase of the simulation. 

As previously explained in section \,\ref{eos}, the iEBE-VISHNU hybrid model transitions from hydrodynamic field equations to microscopic transport at a sudden switching temperature $T_\text{sw}$ 
at which the hydrodynamic energy-momentum tensor is particlized using the Cooper-Frye freezeout prescription,
\begin{equation}
 E\frac{dN_i}{d^3p} = \int_\sigma f_i(x,p) p^\mu d^3\sigma_\mu
 \label{cooper-frye}
\end{equation}
where $f_i$ is the distribution function of particle species $i$, $p^\mu$ is its four-momentum and $d^3\sigma_\mu$ characterizes an element of the isothermal freezeout 
hypersurface defined by $T_\text{sw}$.

The sampled particles then enter the UrQMD simulation where the Boltzmann equation, 
\begin{equation}
 \frac{df_i(x,p)}{dt} = \mathcal{C}_i(x,p),
\end{equation}
is solved to simulate all elastic and inelastic collisions between the particles with collision kernel $\mathcal{C}_i$ until the system becomes too dilute to continue interacting. 
Finally, the four-position, four-momentum and particle identification number of each particle is recorded. 

\begin{figure*}[t]
  \includegraphics[width=\textwidth]{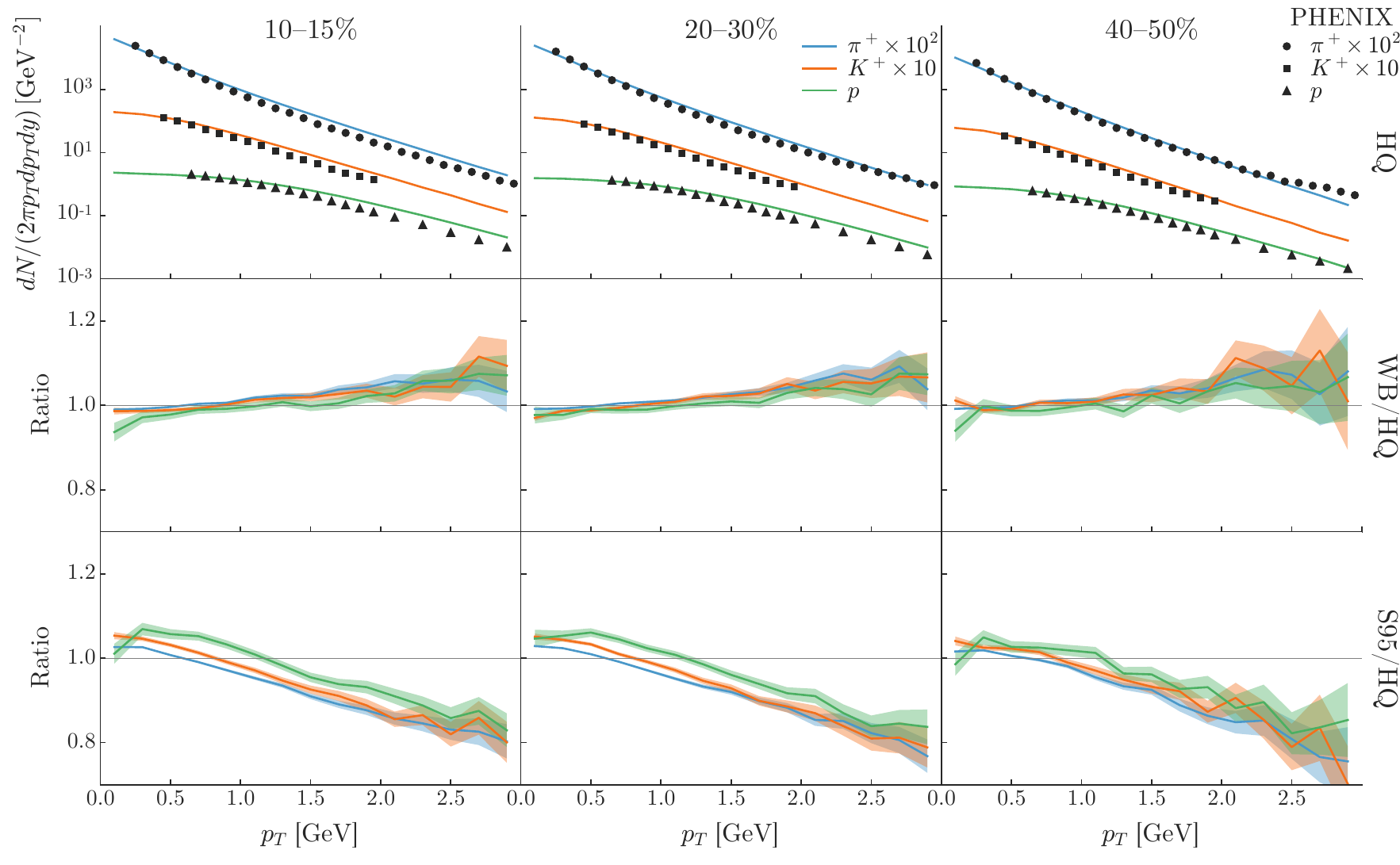}
  \caption{
    \label{fig:spectra} Effect of the equation of state on transverse momentum spectra. Top row: model calculations using the HQ equation of state plotted against 
    PHENIX data \cite{Adler:2003cb} for pions, kaons and protons (blue lines/circles, orange lines/squares and green lines/triangles) in centrality bins 10--15\%, 20--30\% and 40--50\% 
    (columns left to right). Middle and bottom rows: ratios of the WB and S95 invariant yields to the HQ result. Shaded bands indicate two sigma statistical error. }
\end{figure*}

\section{Results}
\label{results}

The results section is organized as follows. In sub-section \ref{spectra} we calculate the particle spectra for each equation of state across three different centrality classes using the final 
particle information output of the hybrid simulation. In sub-section \ref{flow} we repeat the calculation for elliptic and triangular flow but perform the calculation on the 
hydrodynamic Cooper-Frye freezeout surface to reduce statistical errors. In sub-section~\ref{hbt} we calculate the femptoscopic event-averaged Bertsch-Pratt radii, again using the 
final particle information output by the full hybrid calculation. Finally in sub-section \ref{errors}\,, we calculate mean $p_T$ and integrated anisotropic flow cumulants $v_2\{2\}$ and $v_3\{2\}$ from the UrQMD output using a sampling of equation of state curves from the HotQCD published errors. 

All results presented in the following sections are based on $5\cdot10^4$ minimum bias events which are subdivided into centrality classes according to initial entropy, e.g.\ the initial condition events with $20\%$ highest entropy comprise centrality class 0--20\%. Each hydrodynamic event is then oversampled an additional ten times when calculating spectra and flows and
twenty times for pion femptoscopy to suppress finite statistical error.

\subsection{Particle spectra}
\label{spectra}

\begin{figure*}[t]
  \includegraphics[width=\textwidth]{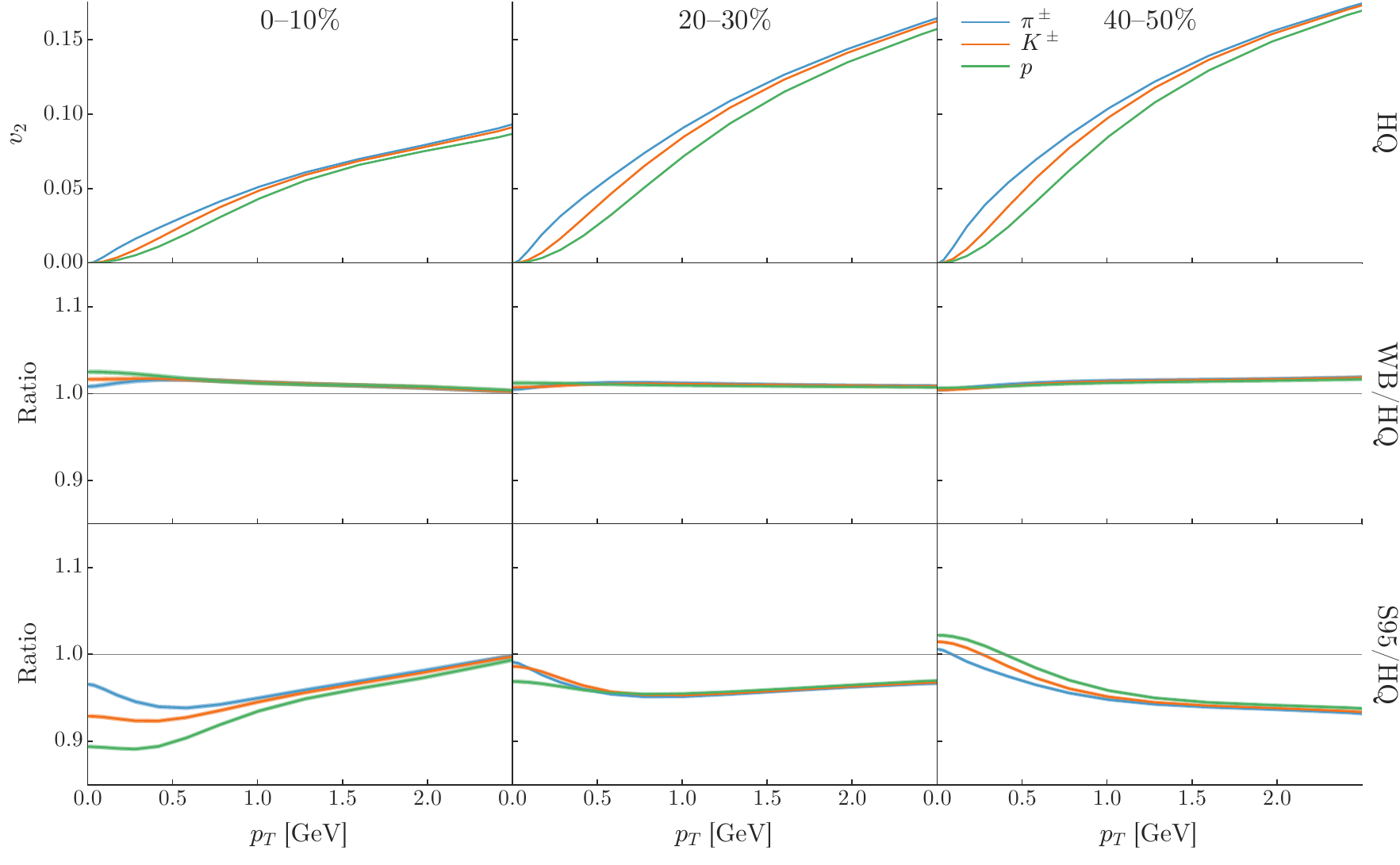}
  \caption{
    \label{fig:v2} Effect of the equation of state on differential elliptic flow $v_2(p_T)$ calculated from the Cooper-Frye freezeout hypersurface \eqref{differential_flow}.
    Top row: model calculations using the HQ equation of state for the elliptic flow $v_2(p_T)$ of pions, kaons and protons (blue, orange and green lines) 
    in centrality bins 0--10\%, 20--30\% and 40--50\% (columns left to right). Middle and bottom rows: ratios of the WB and S95 elliptic flow to 
    the HQ result. Statistical errors are negligible and have been omitted.
  }
\end{figure*}

Figure \ref{fig:spectra} shows the invariant yield $dN/(2\pi p_T dp_T dy)$ of positively charged pions, kaons and protons calculated from the hybrid model for the 10--15\%, 20--30\%
and 40--50\% centrality classes using the HQ, WB and S95 equations of state constructed in section \ref{eos}. 

The first row shows the HQ yields obtained from the hybrid model plotted against observed pion, proton and kaon data from PHENIX \cite{Adler:2003cb}. The second and third rows show the ratio of the invariant yields of the WB and S95 equations of state over the HQ result. One sees that the HQ equation of state provides a good description of observed particle yields except for moderate to large $p_T$ in central collisions where this calculation overpredicts the data. This agreement would likely improve with more realistic initial conditions, bulk viscous corrections and/or more careful treatment of the hydro-to-micro switching temperature $T_\text{sw}$, and thus it's difficult to make any specific statements on the overall fit of model to data. 

The second and third rows of Fig.~\ref{fig:spectra} show the ratios of the WB and S95 yields to the HQ result. The observed spectra predicted by the HQ and WB equations of state agree within statistical error, while the S95 equation of state is appreciably softer and produces $\sim\!5\%$ more particles at $p_T = 0.5$ GeV and $\sim\!20\%$ fewer particles at $p_T=2.5$ GeV across all three centralities.

\vfill

\subsection{Elliptic and triangular flows}
\label{flow}

\begin{figure*}[t]
  \includegraphics[width=\textwidth]{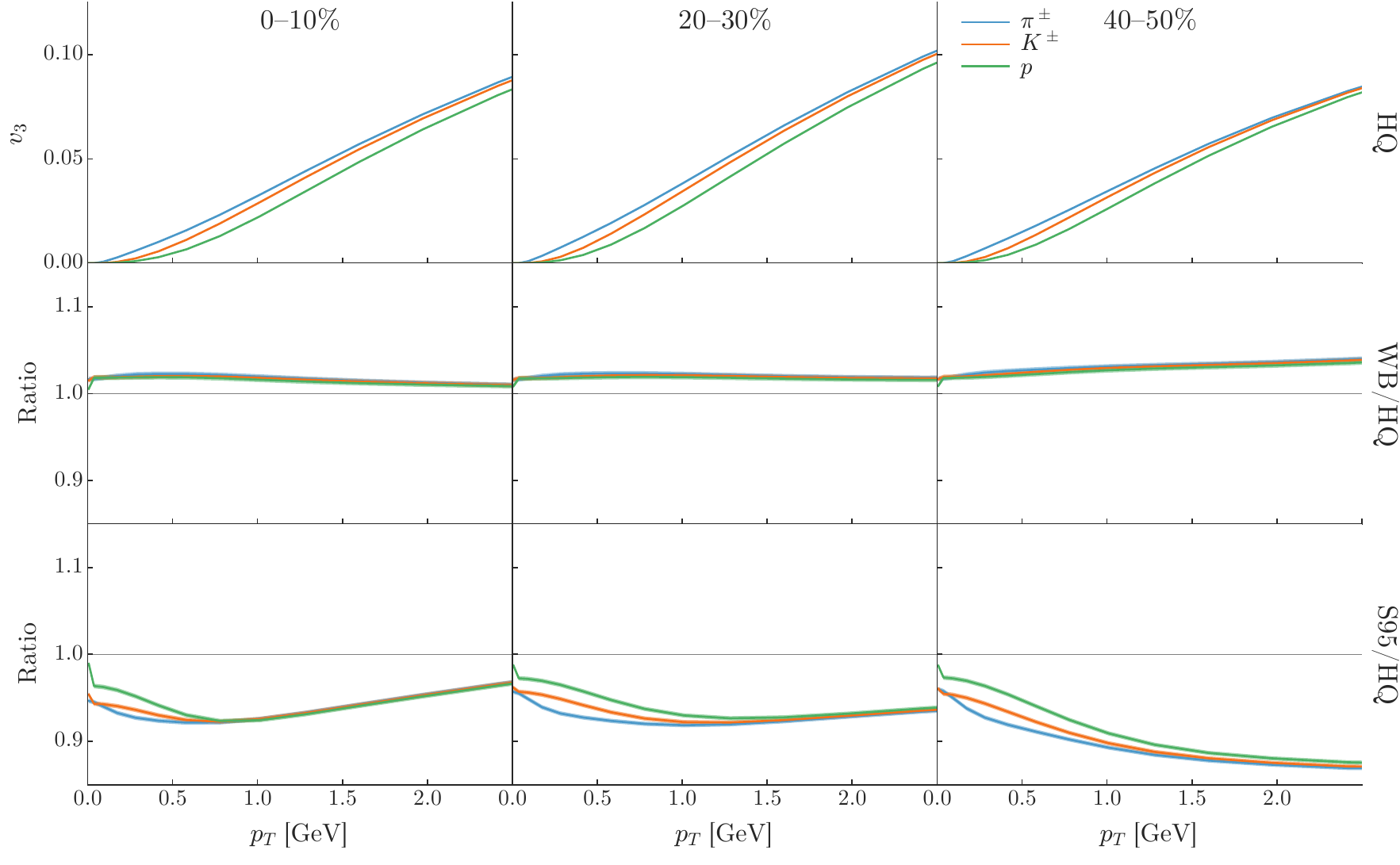}
  \caption{
    \label{fig:v3} Same as Fig.~\ref{fig:v2} but for differential triangular flow $v_3(p_T)$. Note that the y-axis limits in the top row are different.
  }
\end{figure*}

The azimuthal anisotropy of final particle emission is characterized by the Fourier expansion
\begin{equation}
 E \frac{d^3N}{d^3p} = \frac{1}{2\pi} \frac{d^2N}{dy p_T dp_T} \left(1 + \sum\limits_{n=1}^\infty 2 v_n \cos n(\phi - \Psi_{RP}) \right)
\end{equation}
where $\phi$ is the direction of the emitted particle, $\Psi_{RP}$ is the reaction plane angle of the event and $v_n$ the anisotropic flow coefficient corresponding to the Fourier harmonic of order $n$.

The anisotropic flow is typically estimated using multi-particle correlations such as two and four-particle cumulants. The statistical error of the event-averaged estimators is suppressed with both increasing event multiplicity and event sample size. This can pose a challenge for computationally intensive hybrid model calculations which typically cannot reach integrated luminosities comparable to experiment. 

Statistical errors are particularly challenging in differential flow calculations at moderate to large $p_T$ where particle statistics are limited. We circumvent this issue in the differential flow analysis and calculate the flow anisotropy of pions, kaons and protons directly from the Cooper-Frye freezeout surface using the built in routines in the iEBE-VISHNU package according to
\begin{equation}
 \label{differential_flow}
 v_n(p_T) = \frac{\int d\phi_p e^{i n \phi_p} dN/(dy p_T dp_T d\phi_p)}{\int d\phi_p\, dN/(dy p_T dp_T d\phi_p)}.
\end{equation}
Consequently, the flow results in Figs.~\ref{fig:v2} and \ref{fig:v3} do not include contributions from flow generated by the UrQMD hadronic afterburner which is identical for each of the three equations of state.  In Sec.~\ref{errors} results that incorporate UrQMD for the integrated flow measurements will be shown to be consistent.

Figure~\ref{fig:v2} shows the elliptic flow $v_2$ of pions, kaons and protons calculated from equation \eqref{differential_flow} for the HQ, WB and S95 equations of state in 0--10, 20--30 and 40--50\% centrality bins. 
The first row of the figure shows the elliptic flow predicted by the HQ equation of state while the middle and bottom rows display theoretical ratios of the WB and S95 predictions over the HQ result. The presentation
of Fig.~\ref{fig:v3} is identical to Fig.~\ref{fig:v2} except that elliptic flow $v_2$ has been replaced with triangular flow $v_3$.

We see in Fig.~\ref{fig:v2} that the elliptic flow generated by the HQ and WB parameterizations is in very good agreement across all centralities, while the S95 parameterization systematically generates $\sim \! 5\%$ less flow than the HQ equation of state. This is expected as the S95 equation of state is considerably softer in the vicinity of the phase transition as evidenced by the speed of sound in Fig.~\ref{fig:cs}.  In Fig.~\ref{fig:v3}, we see that the effect on the triangular flow is similar to the effect observed on the elliptic flow except more pronounced and generates as large as a $\sim \! 15\%$ discrepancy in the peripheral flows predicted by the HQ and S95 equations of state. 

\begin{figure*}[t]
  \includegraphics[width=\textwidth]{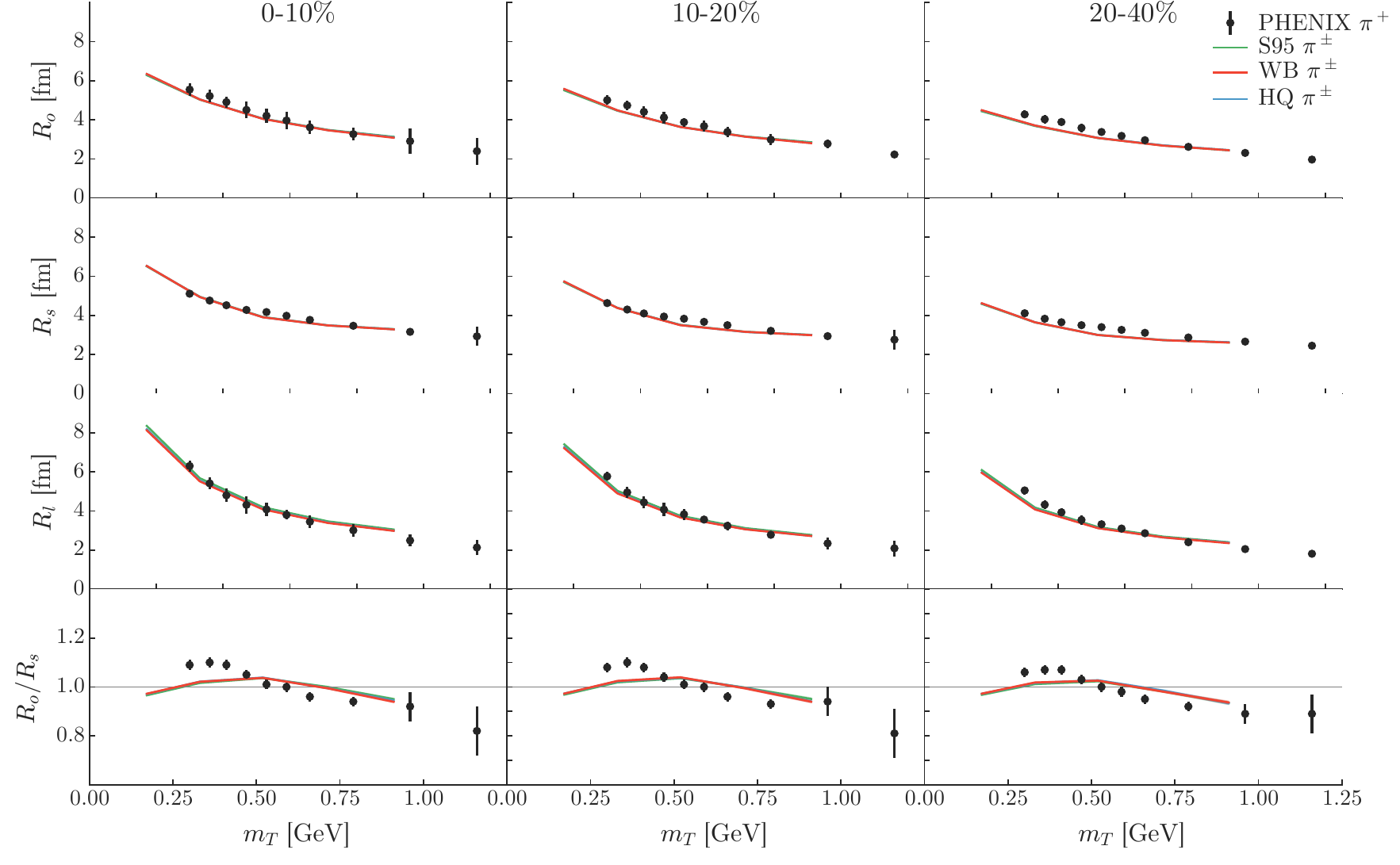}
  \caption{
    \label{fig:hbt} Effect of the equation of state on the Bertsch-Pratt radii. We plot $R_o$, $R_s$, $R_l$ and the ratio $R_o/R_s$ (rows top to bottom) 
    in centrality bins 0--10\%, 10--20\% and 20--40\% (columns left to right) against transverse mass $m_T$ for the HQ, WB and S95 equations of state 
    (blue, red and green lines). Shaded bands indicate two sigma errors from the covariance of the fit parameters \eqref{fitfunction}. Symbols with errors bars 
    are experimental data from PHENIX \cite{Adare:2015bcj}. The HQ, WB and S95 EoS curves overlap and are nearly indistinguishable.
  }
\end{figure*}

\subsection{Femptoscopic Bertsch-Pratt radii}
\label{hbt}

The size of the fireball emission region is obtained using Hanbury-Brown-Twiss (HBT) interferometry for identical particles. The azimuthally averaged two-particle correlation function 
\begin{equation}
 \label{c2}
 C(q, k) = 1 + \frac{\sum\limits_n \sum\limits_{i, j} \delta_{q} \, \delta_{k}\Psi(q,r)}{\sum\limits_{n} \sum\limits_{i,j'} \delta_{q} \, \delta_{k}}
\end{equation}
consists of a numerator with particles pairs sampled from the same event and a denominator with pairs sampled from different events. Here $q = p_i - p_j$ denotes the relative momentum, $r=x_i-x_j$ the relative separation and $k = (p_i + p_j)/2$ the average momentum of the pion pair in the longitudinal co-moving 
frame where the component of $k$ along the beam axis vanishes. The numerator is summed over all events $n$ in a given centrality class and unique particle pair combinations $i,j$ in each event. In the denominator, particle $i$ is
taken from one event and particle $j'$ from a random partner event in the same centrality class. The delta functions $\delta_q$ and $\delta_k$ are $1$ if the momenta $q$ and $k$ fall into their respective bins and $0$ otherwise. Bose-Einstein 
correlations, which are not included natively in the UrQMD model, are imposed by adding the symmetrization factor $\Psi(q,r) = \cos(q\,r)$. 

The average pair momentum $k$ is then projected into its longitudinal component $k_\text{z}$ and transverse component $k_T$, while the separation momentum $q$ is represented in the orthogonal coordinates 
$(q_o, q_s, q_l)$, where $q_l$ lies along the beam axis, $q_o$ is parallel to $k_T$ and $q_s$ perpendicular to $q_o$ and $q_l$. The resulting correlation function is approximated using a Gaussian source and fit to the parametric form
\begin{equation}
 \label{fitfunction}
 C(q_o, q_s, q_l, k_T) = 1 + \lambda\, e^{-(R_o^2 q_o^2 + R_s^2 q_s^2 + R_l^2 q_l^2)}
\end{equation}
using a least squares fit on the three-dimensional correlation function $C(q_o, q_s, q_l)$ to find the optimal source strength $\lambda$ and Bertsch-Pratt radii $R_o$, $R_s$ and $R_l$ for each value of the transverse momentum $k_T$.

We calculate the Bertsch-Pratt radii for each equation of state using identical pions. The fit is performed using $5\cdot10^4$ minimum bias hydrodynamic events and an additional twenty UrQMD oversamples per event. The oversamples are then concatenated into a single particle list to increase the number of particle pairs by a factor of $20^2$.

In Fig.~\ref{fig:hbt}, we plot the Bertsch-Pratt radii for the HQ, WB and S95 equations of state as functions of the transverse mass $m_T = \sqrt{m^2 + k_T^2}$ where $m$ is the pion mass. The horizontal rows show the radii $R_o$, $R_s$, $R_l$ and ratio $R_o/R_s$ (top to bottom), while the columns mark centrality classes 0--10\%, 10--20\% and 20--40\% (left to right). The different colored lines annotated in the legend indicate different equations of state and the bands estimate errors in the fit parameters of Eq. \eqref{fitfunction}. The symbols are experimental data from PHENIX \cite{Adare:2015bcj}.

We see that the Bertsch-Pratt radii predicted by the hybrid model provide a good description of the data across all centralities. However in contrast to spectra and flows, we see no discernible difference in the Bertsch-Pratt radii predicted by the three different equations of state. This suggests that HBT measurements are at most weakly sensitive to small perturbations in the lattice EoS. These results agree with a new sensitivity study which quantified the differential change in simulated observables as a function of perturbed model inputs, e.g.\ the shape of the EoS speed of sound curve \cite{Sangaline:2015isa}. 

\subsection{HotQCD errors}
\label{errors}

\begin{figure}[t]
  \includegraphics[width=\columnwidth]{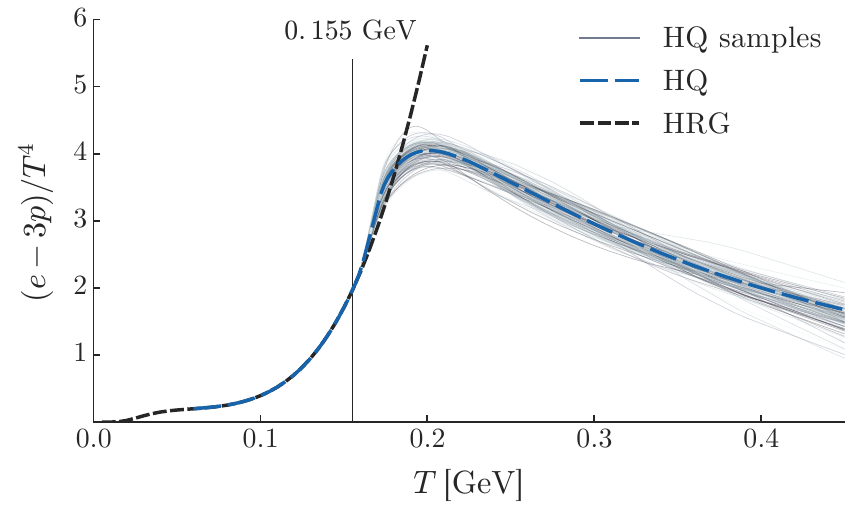}
  \caption{
    \label{fig:splines}
    QCD interaction measure for 100 random samples of the HotQCD error estimate (thin grey lines) plotted alongside the hadron resonance gas EoS (dashed black) and 
    best fit HotQCD (long-dashed blue) parameterization. Both the HQ and HQ sample EoS curves are matched to the HRG EoS at 155 MeV as in Fig.~\ref{fig:trace_final}.
  }
\end{figure}

In addition to the best fit parameterization shown in Fig.~\ref{fig:trace}~, we perform a sensitivity study using equations of state drawn from the HotQCD error distributions.  These curves were calculated in reference \cite{Bazavov:2014pvz} in several steps. The HotQCD trace anomaly was first calculated at various temperatures in the interval $130 < T < 400$ MeV using grids with temporal extent $N_\tau = 8,10$ and $12$.
For each temperature and temporal extent, several thousand lattice configurations were generated, creating a set of ``data points'' with a mean and variance determined from the Monte Carlo ensemble. 
A set of data points was then resampled from the ensemble's mean and variance, and the collection of resampled points, one for each value of the temperature $T$ and grid size $N_\tau$, were fit with the ansatz,
\begin{equation}
 \label{splines}
 \frac{\theta^{\mu\mu}(T)}{T^4} = A + \sum\limits_{i=1}^{n_k=3} B_i \times S_i(T) + \frac{C + \sum_{i=1}^{n_k + 3} D_i \times S_i(T)}{N_\tau^2}.
\end{equation}
Here the constants $A$, $B_i$, $C$ and $D_i$ are parameters of the fit, $S_i$ is a set of cubic basis splines and $n_k$ the number of knots used in the B-spline fitting. 
The entire procedure was repeated 20,001 times to sample the function space of $\theta^{\mu\mu}(T)/T^4$ from the errors in the ensemble averaged lattice measurements.

Here we investigate the effect of these HotQCD lattice errors by measuring the spectra and flows for a subset of 100 randomly sampled EoS curves determined according to equation \eqref{splines}.
The piecewise interpolation procedure described in section \ref{eos} is applied to each spline to smoothly match the HotQCD lattice interaction measures with the HRG result at low temperature. 
The resulting interaction measures are shown in Fig.~\ref{fig:splines} alongside the HRG-matched best fit HotQCD parameterization which naturally falls in the middle of the sampled curves.  

The energy density, entropy density, pressure and temperature are then calculated from each interaction measure according to \eqref{conversion} to generate 100 different EoS tables. 
Above 400~MeV, the higher derivatives of the interaction measures become unreliable, and we extrapolate the EoS table using a simple power law, e.g.\ the energy density as a function of temperature is extended using $e(T) = a T^b$ where the coefficients $a$ and $b$ are tuned to fit the lattice EoS at 400 MeV. We note, however, that this modification has negligible impact on the hydrodynamic evolution at RHIC where the spacetime volume of the system is predominantly below 400~MeV.

In the previous sections, we compared observables calculated from different EoS both as functions of transverse momentum and centrality, as well as for different particle species. Figures~\ref{fig:spectra}--\ref{fig:hbt} indicate that changes in the EoS affect pions, kaons and protons in a similar fashion. Meanwhile, the $p_T$ dependence of these quantities exhibits a few general trends. Changing the stiffness of the EoS changes the slope of the spectra while it shifts the differential flow curves vertically up and down. These generic features suggest that simpler quantities such as mean $p_T$ and integrated flow may offer equal resolving power to species dependent and differential quantities with the added benefit of increased statistics and reduced model uncertainty.

\begin{figure}
  \includegraphics[width=\columnwidth]{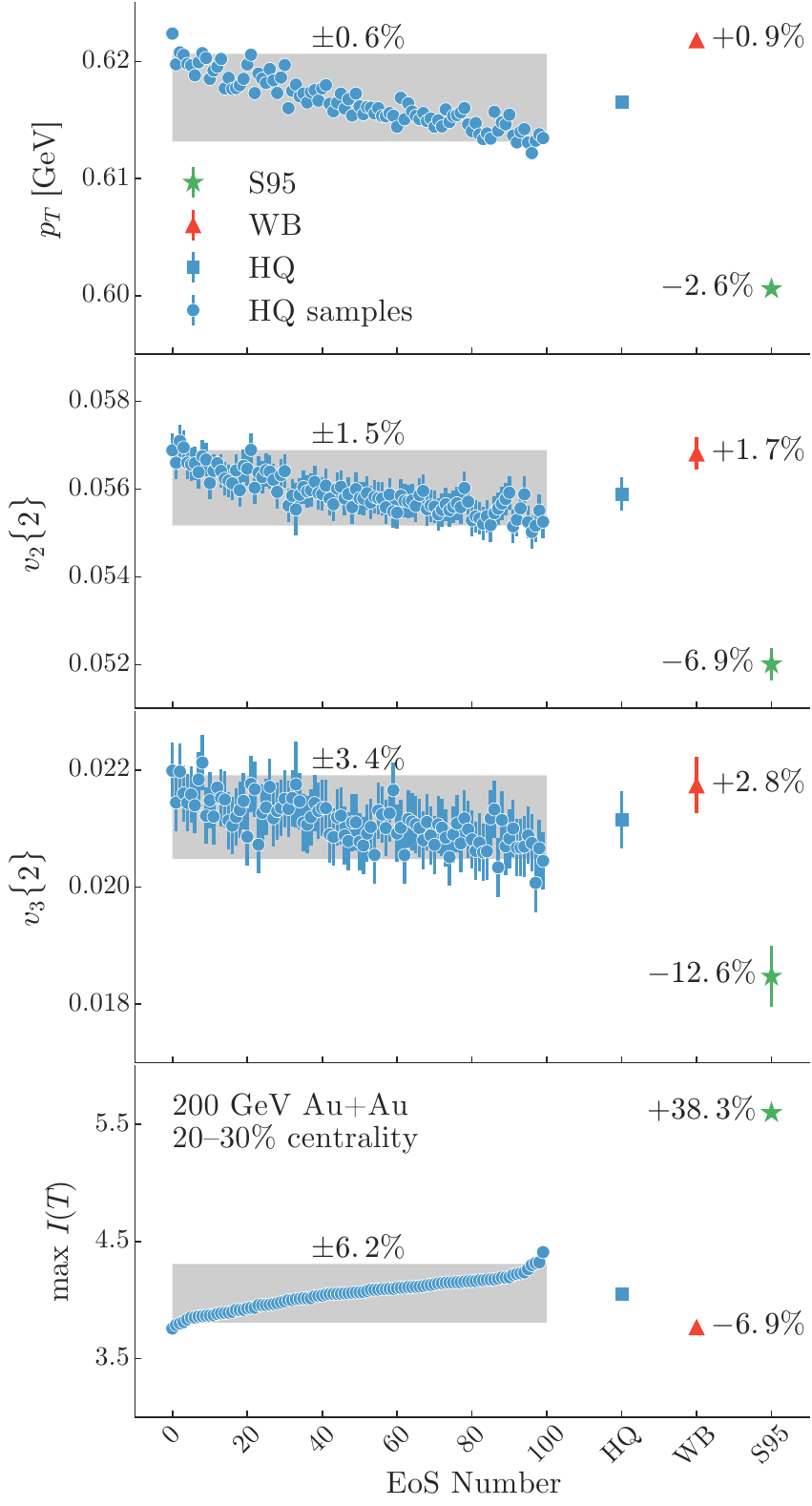}
  \caption{
    \label{fig:eos_compare}
    From top to bottom: calculations of mean $p_T$, elliptic flow cumulant $v_2\{2\}$, triangular flow cumulant $v_3\{2\}$ and maximum value of the interaction measure 
    $I(T) = (e - 3 p)/T^4$ for $\sqrt{s_{NN}}=200$ GeV Au+Au collisions in centrality bin 20--30\%. The calculation is performed for 100 EoS curves randomly sampled from the 
    errors in the HotQCD continuum extrapolation (blue circles) as well as for the HQ, WB and S95 EoS curves shown in Fig.~\ref{fig:trace_final} (blue squares, red triangles 
    and green stars). Gray shaded bands show the two-sigma confidence interval for the HQ samples and percents indicate the relative increase (decrease) of a given observable for each EoS 
    relative to the HQ result (blue square). HQ EoS samples are numbered in increasing order by the peak value of the interaction measure. The kinematic cuts are $p_T < 3$ GeV with 
    $|\eta| < 0.5$ for mean $p_T$ and $|\eta| < 1$ for flows. Vertical error bars on the measurements represent two-sigma statistical error from finite particle fluctuations.
  }
\end{figure}

With this in mind, we quantify the effect of the HotQCD errors by calculating the mean $p_T$ and integrated two-particle cumulants $v_2\{2\}$ and $v_3\{2\}$ for all charged particles in the 20--30\% centrality bin. Unlike the $p_T$-differential flows in Figs.~\ref{fig:v2} and \ref{fig:v3}, these integrated cumulants are calculated from the UrQMD particle output and account for flow developed in the hadronic phase of the collision. 

The mean $p_T$ and flow cumulants for the sampled HotQCD EoS curves--numbered in increasing order by the maximum value of their respective interaction measures--are displayed Fig.~\ref{fig:eos_compare} alongside results for the HQ, WB and S95 EoS described in section \ref{eos}. The grey band plotted on top of the HQ samples marks the two-sigma confidence interval describing 95\% of the variance in the HQ samples while the percentages next to the data points describe the increase (decrease) of each EoS relative to the HQ EoS result (blue square). For comparison, the bottom panel of Fig.~\ref{fig:eos_compare} displays the maximum value of the interaction measure for each EoS.

Several key features are immediately apparent from the figure. We see a clear separation of the different EoS curves which is strongly correlated with the maximum value of the interaction measure (bottom panel). Softer EoS curves have a larger peak in the trace anomaly and hence drive less radial, elliptic and triangular flow as evidenced by the smaller values of 
mean $p_T$, $v_2$ and $v_3$.

Errors in the HotQCD continuum extrapolation, represented by the spread in the HQ EoS samples (blue circles), account for small (order 1\%) differences in mean $p_T$, $v_2$ and $v_3$ which are similar in magnitude to differences between the Wuppertal-Budapest stout fermion (red triangles) and HotQCD HISQ/tree actions (blue squares). On the other hand, the pronounced peak in the S95 interaction measure leads to much larger differences in mean $p_T$ and flows. For example, the value of $v_3\{2\}$ calculated using the S95 EoS is 12.6\% smaller than the same calculation performed with the HQ EoS.

\section{Conclusion and Outlook}
\label{conclusion}

The LQCD EoS is an essential ingredient used in hydrodynamic simulations of relativistic heavy-ion collisions. In this study, we simulated collisions at RHIC using a modern event-by-event hybrid model with several calculations of the LQCD equation of state to quantify differences in the simulated spectra, flow and HBT radii. 

The analysis was performed in two stages. In the first stage of the analysis, we compared simulation results obtained with state-of-the-art LQCD EoS calculations from the HotQCD collaboration using the HISQ/tree fermion action and Wuppertal-Budapest collaboration using the stout fermion action, as well as using the older s95p-v1 parameterization constructed from coarser lattices using the p4 action without continuum extrapolation. The three parameterizations are each matched to a hadron resonance gas EoS at $T=155$ MeV where the hybrid model transitions from viscous relativistic fluid dynamics to Boltzmann transport described by the UrQMD model. For each EoS, we calculate spectra, differential flows and HBT radii for pions, kaons and protons using three different centrality classes.

We find that the spectra and flows of the HotQCD and Wuppertal-Budapest calculations are largely indistinguishable, while the s95p-v1 parameterization leads to noticeably softer spectra and less anisotropic flow. On the other hand, measurements of the azimuthally averaged HBT radii were not sensitive enough to resolve differences between the different EoS parameterizations. Furthermore, we see little differences for pions, kaons or protons and somewhat surprisingly only moderate sensitivity of the EoS deviations to changes in the centrality class.

In the second stage of the analysis, we quantified the effect of errors in the HotQCD continuum extrapolation using a set of 100 randomly sampled EoS curves from the HotQCD error estimate. The mean $p_T$ and integrated flow cumulants $v_2\{2\}$ and $v_3\{2\}$ were calculated for each of the HotQCD EoS samples as well as for the HotQCD, Wuppertal-Budapest and s95p-v1 EoS parameterizations used in the first stage of our analysis. We observe that errors in the HotQCD continuum extrapolation lead to less than 1\% differences in mean $p_T$ and 2--3\% variations in $v_2\{2\}$ and $v_3\{2\}$. These errors are comparable in magnitude to current experimental systematic errors~\cite{Adamczyk:2013jh, Adare:2011uo}, and until significant error reductions are obtained to constrain other model input parameters, further refinements of the equation of state at zero baryon density are unlikely to be needed. However, continued use of the s95p-v1 equation of state in hydrodynamic modeling will produce particle spectra that are too soft and $v_2$ and $v_3$ values that are order $10\%$ too small.

\begin{acknowledgments}
We wish to thank Jonah Bernard, Ulrich Heinz and Christopher Plumberg for many helpful discussions.
This work was performed with support from the U.S. Department of Energy NNSA Stockpile Stewardship Graduate Fellowship under grant no.~DE-FC52-08NA28752 and from Lawrence Livermore National Laboratory under Contract DE-AC52-07NA27344.
\end{acknowledgments}

\bibliography{eos,duke-qcd-refs/Duke_QCD_refs}

\begin{thebibliography}{34}%
\makeatletter
\providecommand \@ifxundefined [1]{%
 \@ifx{#1\undefined}
}%
\providecommand \@ifnum [1]{%
 \ifnum #1\expandafter \@firstoftwo
 \else \expandafter \@secondoftwo
 \fi
}%
\providecommand \@ifx [1]{%
 \ifx #1\expandafter \@firstoftwo
 \else \expandafter \@secondoftwo
 \fi
}%
\providecommand \natexlab [1]{#1}%
\providecommand \enquote  [1]{``#1''}%
\providecommand \bibnamefont  [1]{#1}%
\providecommand \bibfnamefont [1]{#1}%
\providecommand \citenamefont [1]{#1}%
\providecommand \href@noop [0]{\@secondoftwo}%
\providecommand \href [0]{\begingroup \@sanitize@url \@href}%
\providecommand \@href[1]{\@@startlink{#1}\@@href}%
\providecommand \@@href[1]{\endgroup#1\@@endlink}%
\providecommand \@sanitize@url [0]{\catcode `\\12\catcode `\$12\catcode
  `\&12\catcode `\#12\catcode `\^12\catcode `\_12\catcode `\%12\relax}%
\providecommand \@@startlink[1]{}%
\providecommand \@@endlink[0]{}%
\providecommand \url  [0]{\begingroup\@sanitize@url \@url }%
\providecommand \@url [1]{\endgroup\@href {#1}{\urlprefix }}%
\providecommand \urlprefix  [0]{URL }%
\providecommand \Eprint [0]{\href }%
\providecommand \doibase [0]{http://dx.doi.org/}%
\providecommand \selectlanguage [0]{\@gobble}%
\providecommand \bibinfo  [0]{\@secondoftwo}%
\providecommand \bibfield  [0]{\@secondoftwo}%
\providecommand \translation [1]{[#1]}%
\providecommand \BibitemOpen [0]{}%
\providecommand \bibitemStop [0]{}%
\providecommand \bibitemNoStop [0]{.\EOS\space}%
\providecommand \EOS [0]{\spacefactor3000\relax}%
\providecommand \BibitemShut  [1]{\csname bibitem#1\endcsname}%
\let\auto@bib@innerbib\@empty
\bibitem [{\citenamefont {Huovinen}(2005)}]{Huovinen:2005gy}%
  \BibitemOpen
  \bibfield  {author} {\bibinfo {author} {\bibfnamefont {P.}~\bibnamefont
  {Huovinen}},\ }\href {\doibase 10.1016/j.nuclphysa.2005.07.016} {\bibfield
  {journal} {\bibinfo  {journal} {Nucl. Phys.}\ }\textbf {\bibinfo {volume}
  {A761}},\ \bibinfo {pages} {296} (\bibinfo {year} {2005})},\ \Eprint
  {http://arxiv.org/abs/nucl-th/0505036} {arXiv:nucl-th/0505036 [nucl-th]}
  \BibitemShut {NoStop}%
\bibitem [{\citenamefont {Huovinen}\ and\ \citenamefont
  {Petreczky}(2010)}]{Huovinen:2009yb}%
  \BibitemOpen
  \bibfield  {author} {\bibinfo {author} {\bibfnamefont {P.}~\bibnamefont
  {Huovinen}}\ and\ \bibinfo {author} {\bibfnamefont {P.}~\bibnamefont
  {Petreczky}},\ }\href {\doibase 10.1016/j.nuclphysa.2010.02.015} {\bibfield
  {journal} {\bibinfo  {journal} {Nucl. Phys.}\ }\textbf {\bibinfo {volume}
  {A837}},\ \bibinfo {pages} {26} (\bibinfo {year} {2010})},\ \Eprint
  {http://arxiv.org/abs/0912.2541} {arXiv:0912.2541 [hep-ph]} \BibitemShut
  {NoStop}%
\bibitem [{\citenamefont {Pratt}\ \emph {et~al.}(2015)\citenamefont {Pratt},
  \citenamefont {Sangaline}, \citenamefont {Sorensen},\ and\ \citenamefont
  {Wang}}]{Pratt:2015zsa}%
  \BibitemOpen
  \bibfield  {author} {\bibinfo {author} {\bibfnamefont {S.}~\bibnamefont
  {Pratt}}, \bibinfo {author} {\bibfnamefont {E.}~\bibnamefont {Sangaline}},
  \bibinfo {author} {\bibfnamefont {P.}~\bibnamefont {Sorensen}}, \ and\
  \bibinfo {author} {\bibfnamefont {H.}~\bibnamefont {Wang}},\ }\href {\doibase
  10.1103/PhysRevLett.114.202301} {\bibfield  {journal} {\bibinfo  {journal}
  {Phys. Rev. Lett.}\ }\textbf {\bibinfo {volume} {114}},\ \bibinfo {pages}
  {202301} (\bibinfo {year} {2015})},\ \Eprint
  {http://arxiv.org/abs/1501.04042} {arXiv:1501.04042 [nucl-th]} \BibitemShut
  {NoStop}%
\bibitem [{\citenamefont {Sangaline}\ and\ \citenamefont
  {Pratt}(2015)}]{Sangaline:2015isa}%
  \BibitemOpen
  \bibfield  {author} {\bibinfo {author} {\bibfnamefont {E.}~\bibnamefont
  {Sangaline}}\ and\ \bibinfo {author} {\bibfnamefont {S.}~\bibnamefont
  {Pratt}},\ }\href@noop {} {\  (\bibinfo {year} {2015})},\ \Eprint
  {http://arxiv.org/abs/1508.07017} {arXiv:1508.07017 [nucl-th]} \BibitemShut
  {NoStop}%
\bibitem [{\citenamefont {Borsanyi}\ \emph {et~al.}(2014)\citenamefont
  {Borsanyi}, \citenamefont {Fodor}, \citenamefont {Hoelbling}, \citenamefont
  {Katz}, \citenamefont {Krieg},\ and\ \citenamefont
  {Szabo}}]{Borsanyi:2013bia}%
  \BibitemOpen
  \bibfield  {author} {\bibinfo {author} {\bibfnamefont {S.}~\bibnamefont
  {Borsanyi}}, \bibinfo {author} {\bibfnamefont {Z.}~\bibnamefont {Fodor}},
  \bibinfo {author} {\bibfnamefont {C.}~\bibnamefont {Hoelbling}}, \bibinfo
  {author} {\bibfnamefont {S.~D.}\ \bibnamefont {Katz}}, \bibinfo {author}
  {\bibfnamefont {S.}~\bibnamefont {Krieg}}, \ and\ \bibinfo {author}
  {\bibfnamefont {K.~K.}\ \bibnamefont {Szabo}},\ }\href {\doibase
  10.1016/j.physletb.2014.01.007} {\bibfield  {journal} {\bibinfo  {journal}
  {Phys. Lett.}\ }\textbf {\bibinfo {volume} {B730}},\ \bibinfo {pages} {99}
  (\bibinfo {year} {2014})},\ \Eprint {http://arxiv.org/abs/1309.5258}
  {arXiv:1309.5258 [hep-lat]} \BibitemShut {NoStop}%
\bibitem [{\citenamefont {Bazavov}\ \emph
  {et~al.}(2014{\natexlab{a}})\citenamefont {Bazavov} \emph
  {et~al.}}]{Bazavov:2014pvz}%
  \BibitemOpen
  \bibfield  {author} {\bibinfo {author} {\bibfnamefont {A.}~\bibnamefont
  {Bazavov}} \emph {et~al.} (\bibinfo {collaboration} {HotQCD}),\ }\href
  {\doibase 10.1103/PhysRevD.90.094503} {\bibfield  {journal} {\bibinfo
  {journal} {Phys. Rev.}\ }\textbf {\bibinfo {volume} {D90}},\ \bibinfo {pages}
  {094503} (\bibinfo {year} {2014}{\natexlab{a}})},\ \Eprint
  {http://arxiv.org/abs/1407.6387} {arXiv:1407.6387 [hep-lat]} \BibitemShut
  {NoStop}%
\bibitem [{\citenamefont {Bazavov}\ \emph {et~al.}(2009)\citenamefont {Bazavov}
  \emph {et~al.}}]{Bazavov:2009zn}%
  \BibitemOpen
  \bibfield  {author} {\bibinfo {author} {\bibfnamefont {A.}~\bibnamefont
  {Bazavov}} \emph {et~al.},\ }\href {\doibase 10.1103/PhysRevD.80.014504}
  {\bibfield  {journal} {\bibinfo  {journal} {Phys. Rev.}\ }\textbf {\bibinfo
  {volume} {D80}},\ \bibinfo {pages} {014504} (\bibinfo {year} {2009})},\
  \Eprint {http://arxiv.org/abs/0903.4379} {arXiv:0903.4379 [hep-lat]}
  \BibitemShut {NoStop}%
\bibitem [{\citenamefont {Soltz}\ \emph {et~al.}(2015)\citenamefont {Soltz},
  \citenamefont {DeTar}, \citenamefont {Karsch}, \citenamefont {Mukherjee},\
  and\ \citenamefont {Vranas}}]{Soltz:2015ula}%
  \BibitemOpen
  \bibfield  {author} {\bibinfo {author} {\bibfnamefont {R.~A.}\ \bibnamefont
  {Soltz}}, \bibinfo {author} {\bibfnamefont {C.}~\bibnamefont {DeTar}},
  \bibinfo {author} {\bibfnamefont {F.}~\bibnamefont {Karsch}}, \bibinfo
  {author} {\bibfnamefont {S.}~\bibnamefont {Mukherjee}}, \ and\ \bibinfo
  {author} {\bibfnamefont {P.}~\bibnamefont {Vranas}},\ }\href@noop {}
  {\bibfield  {journal} {\bibinfo  {journal} {Ann. Rev. Nucl. Part. Sci.}\
  }\textbf {\bibinfo {volume} {65}},\ \bibinfo {pages} {379} (\bibinfo {year}
  {2015})},\ \Eprint {http://arxiv.org/abs/1502.02296} {arXiv:1502.02296
  [hep-lat]} \BibitemShut {NoStop}%
\bibitem [{\citenamefont {Bass}\ \emph {et~al.}(1998)\citenamefont {Bass} \emph
  {et~al.}}]{Bass:1998ca}%
  \BibitemOpen
  \bibfield  {author} {\bibinfo {author} {\bibfnamefont {S.~A.}\ \bibnamefont
  {Bass}} \emph {et~al.},\ }\href {\doibase 10.1016/S0146-6410(98)00058-1}
  {\bibfield  {journal} {\bibinfo  {journal} {Prog. Part. Nucl. Phys.}\
  }\textbf {\bibinfo {volume} {41}},\ \bibinfo {pages} {255} (\bibinfo {year}
  {1998})},\ \bibinfo {note} {[Prog. Part. Nucl. Phys.41,225(1998)]},\ \Eprint
  {http://arxiv.org/abs/nucl-th/9803035} {arXiv:nucl-th/9803035 [nucl-th]}
  \BibitemShut {NoStop}%
\bibitem [{\citenamefont {Bleicher}\ \emph {et~al.}(1999)\citenamefont
  {Bleicher} \emph {et~al.}}]{Bleicher:1999xi}%
  \BibitemOpen
  \bibfield  {author} {\bibinfo {author} {\bibfnamefont {M.}~\bibnamefont
  {Bleicher}} \emph {et~al.},\ }\href {\doibase 10.1088/0954-3899/25/9/308}
  {\bibfield  {journal} {\bibinfo  {journal} {J. Phys.}\ }\textbf {\bibinfo
  {volume} {G25}},\ \bibinfo {pages} {1859} (\bibinfo {year} {1999})},\ \Eprint
  {http://arxiv.org/abs/hep-ph/9909407} {arXiv:hep-ph/9909407 [hep-ph]}
  \BibitemShut {NoStop}%
\bibitem [{\citenamefont {Bazavov}\ \emph
  {et~al.}(2014{\natexlab{b}})\citenamefont {Bazavov} \emph
  {et~al.}}]{Bazavov:2014xya}%
  \BibitemOpen
  \bibfield  {author} {\bibinfo {author} {\bibfnamefont {A.}~\bibnamefont
  {Bazavov}} \emph {et~al.},\ }\href {\doibase 10.1103/PhysRevLett.113.072001}
  {\bibfield  {journal} {\bibinfo  {journal} {Phys. Rev. Lett.}\ }\textbf
  {\bibinfo {volume} {113}},\ \bibinfo {pages} {072001} (\bibinfo {year}
  {2014}{\natexlab{b}})},\ \Eprint {http://arxiv.org/abs/1404.6511}
  {arXiv:1404.6511 [hep-lat]} \BibitemShut {NoStop}%
\bibitem [{\citenamefont {Adare}\ \emph
  {et~al.}(2015{\natexlab{a}})\citenamefont {Adare} \emph
  {et~al.}}]{Adare:2015aqk}%
  \BibitemOpen
  \bibfield  {author} {\bibinfo {author} {\bibfnamefont {A.}~\bibnamefont
  {Adare}} \emph {et~al.} (\bibinfo {collaboration} {PHENIX}),\ }\href@noop {}
  {\  (\bibinfo {year} {2015}{\natexlab{a}})},\ \Eprint
  {http://arxiv.org/abs/1506.07834} {arXiv:1506.07834 [nucl-ex]} \BibitemShut
  {NoStop}%
\bibitem [{\citenamefont {Bors{\'a}nyi}\ \emph {et~al.}(2010)\citenamefont
  {Bors{\'a}nyi}, \citenamefont {Fodor}, \citenamefont {Hoelbling},
  \citenamefont {Katz}, \citenamefont {Krieg}, \citenamefont {Ratti},\ and\
  \citenamefont {Szab{\'o}}}]{Borsanyi:2010gc}%
  \BibitemOpen
  \bibfield  {author} {\bibinfo {author} {\bibfnamefont {S.}~\bibnamefont
  {Bors{\'a}nyi}}, \bibinfo {author} {\bibfnamefont {Z.}~\bibnamefont {Fodor}},
  \bibinfo {author} {\bibfnamefont {C.}~\bibnamefont {Hoelbling}}, \bibinfo
  {author} {\bibfnamefont {S.~D.}\ \bibnamefont {Katz}}, \bibinfo {author}
  {\bibfnamefont {S.}~\bibnamefont {Krieg}}, \bibinfo {author} {\bibfnamefont
  {C.}~\bibnamefont {Ratti}}, \ and\ \bibinfo {author} {\bibfnamefont {K.~K.}\
  \bibnamefont {Szab{\'o}}},\ }\href@noop {} {\bibfield  {journal} {\bibinfo
  {journal} {Journal of High Energy Physics}\ }\textbf {\bibinfo {volume}
  {2010}} (\bibinfo {year} {2010})}\BibitemShut {NoStop}%
\bibitem [{\citenamefont {Bazavov}\ \emph {et~al.}(2012)\citenamefont {Bazavov}
  \emph {et~al.}}]{Bazavov:2012iu}%
  \BibitemOpen
  \bibfield  {author} {\bibinfo {author} {\bibfnamefont {A.}~\bibnamefont
  {Bazavov}} \emph {et~al.} (\bibinfo {collaboration} {HotQCD}),\ }\href@noop
  {} {\bibfield  {journal} {\bibinfo  {journal} {Phys. Rev. D}\ }\textbf
  {\bibinfo {volume} {85}},\ \bibinfo {pages} {054503} (\bibinfo {year}
  {2012})}\BibitemShut {NoStop}%
\bibitem [{\citenamefont {Bhattacharya}\ \emph {et~al.}(2014)\citenamefont
  {Bhattacharya} \emph {et~al.}}]{Bhattacharya:2014iw}%
  \BibitemOpen
  \bibfield  {author} {\bibinfo {author} {\bibfnamefont {T.}~\bibnamefont
  {Bhattacharya}} \emph {et~al.} (\bibinfo {collaboration} {HotQCD}),\
  }\href@noop {} {\bibfield  {journal} {\bibinfo  {journal} {Phys. Rev. Lett.}\
  }\textbf {\bibinfo {volume} {113}},\ \bibinfo {pages} {082001} (\bibinfo
  {year} {2014})}\BibitemShut {NoStop}%
\bibitem [{\citenamefont {Song}\ and\ \citenamefont
  {Heinz}(2008)}]{Song:2007ux}%
  \BibitemOpen
  \bibfield  {author} {\bibinfo {author} {\bibfnamefont {H.}~\bibnamefont
  {Song}}\ and\ \bibinfo {author} {\bibfnamefont {U.~W.}\ \bibnamefont
  {Heinz}},\ }\href {\doibase 10.1103/PhysRevC.77.064901} {\bibfield  {journal}
  {\bibinfo  {journal} {Phys. Rev.}\ }\textbf {\bibinfo {volume} {C77}},\
  \bibinfo {pages} {064901} (\bibinfo {year} {2008})},\ \Eprint
  {http://arxiv.org/abs/0712.3715} {arXiv:0712.3715 [nucl-th]} \BibitemShut
  {NoStop}%
\bibitem [{\citenamefont {Shen}\ \emph {et~al.}(2014)\citenamefont {Shen},
  \citenamefont {Qiu}, \citenamefont {Song}, \citenamefont {Bernhard},
  \citenamefont {Bass},\ and\ \citenamefont {Heinz}}]{Shen:2014vra}%
  \BibitemOpen
  \bibfield  {author} {\bibinfo {author} {\bibfnamefont {C.}~\bibnamefont
  {Shen}}, \bibinfo {author} {\bibfnamefont {Z.}~\bibnamefont {Qiu}}, \bibinfo
  {author} {\bibfnamefont {H.}~\bibnamefont {Song}}, \bibinfo {author}
  {\bibfnamefont {J.}~\bibnamefont {Bernhard}}, \bibinfo {author}
  {\bibfnamefont {S.}~\bibnamefont {Bass}}, \ and\ \bibinfo {author}
  {\bibfnamefont {U.}~\bibnamefont {Heinz}},\ }\href@noop {} {\  (\bibinfo
  {year} {2014})},\ \Eprint {http://arxiv.org/abs/1409.8164} {arXiv:1409.8164
  [nucl-th]} \BibitemShut {NoStop}%
\bibitem [{\citenamefont {Schenke}\ \emph {et~al.}(2012)\citenamefont
  {Schenke}, \citenamefont {Tribedy},\ and\ \citenamefont
  {Venugopalan}}]{Schenke:2012wb}%
  \BibitemOpen
  \bibfield  {author} {\bibinfo {author} {\bibfnamefont {B.}~\bibnamefont
  {Schenke}}, \bibinfo {author} {\bibfnamefont {P.}~\bibnamefont {Tribedy}}, \
  and\ \bibinfo {author} {\bibfnamefont {R.}~\bibnamefont {Venugopalan}},\
  }\href {\doibase 10.1103/PhysRevLett.108.252301} {\bibfield  {journal}
  {\bibinfo  {journal} {Phys. Rev. Lett.}\ }\textbf {\bibinfo {volume} {108}},\
  \bibinfo {pages} {252301} (\bibinfo {year} {2012})},\ \Eprint
  {http://arxiv.org/abs/1202.6646} {arXiv:1202.6646 [nucl-th]} \BibitemShut
  {NoStop}%
\bibitem [{\citenamefont {Niemi}\ \emph {et~al.}(2015)\citenamefont {Niemi},
  \citenamefont {Eskola},\ and\ \citenamefont {Paatelainen}}]{Niemi:2015qia}%
  \BibitemOpen
  \bibfield  {author} {\bibinfo {author} {\bibfnamefont {H.}~\bibnamefont
  {Niemi}}, \bibinfo {author} {\bibfnamefont {K.~J.}\ \bibnamefont {Eskola}}, \
  and\ \bibinfo {author} {\bibfnamefont {R.}~\bibnamefont {Paatelainen}},\
  }\href@noop {} {\  (\bibinfo {year} {2015})},\ \Eprint
  {http://arxiv.org/abs/1505.02677} {arXiv:1505.02677 [hep-ph]} \BibitemShut
  {NoStop}%
\bibitem [{\citenamefont {Chatterjee}\ \emph {et~al.}(2015)\citenamefont
  {Chatterjee}, \citenamefont {Singh}, \citenamefont {Ghosh}, \citenamefont
  {Hasanujjaman}, \citenamefont {Alam},\ and\ \citenamefont
  {Sarkar}}]{Chatterjee:2015aja}%
  \BibitemOpen
  \bibfield  {author} {\bibinfo {author} {\bibfnamefont {S.}~\bibnamefont
  {Chatterjee}}, \bibinfo {author} {\bibfnamefont {S.~K.}\ \bibnamefont
  {Singh}}, \bibinfo {author} {\bibfnamefont {S.}~\bibnamefont {Ghosh}},
  \bibinfo {author} {\bibfnamefont {M.}~\bibnamefont {Hasanujjaman}}, \bibinfo
  {author} {\bibfnamefont {J.}~\bibnamefont {Alam}}, \ and\ \bibinfo {author}
  {\bibfnamefont {S.}~\bibnamefont {Sarkar}},\ }\href@noop {} {\  (\bibinfo
  {year} {2015})},\ \Eprint {http://arxiv.org/abs/1510.01311} {arXiv:1510.01311
  [nucl-th]} \BibitemShut {NoStop}%
\bibitem [{\citenamefont {Moreland}\ \emph {et~al.}(2015)\citenamefont
  {Moreland}, \citenamefont {Bernhard},\ and\ \citenamefont
  {Bass}}]{Moreland:2014oya}%
  \BibitemOpen
  \bibfield  {author} {\bibinfo {author} {\bibfnamefont {J.~S.}\ \bibnamefont
  {Moreland}}, \bibinfo {author} {\bibfnamefont {J.~E.}\ \bibnamefont
  {Bernhard}}, \ and\ \bibinfo {author} {\bibfnamefont {S.~A.}\ \bibnamefont
  {Bass}},\ }\href {\doibase 10.1103/PhysRevC.92.011901} {\bibfield  {journal}
  {\bibinfo  {journal} {Phys. Rev.}\ }\textbf {\bibinfo {volume} {C92}},\
  \bibinfo {pages} {011901} (\bibinfo {year} {2015})},\ \Eprint
  {http://arxiv.org/abs/1412.4708} {arXiv:1412.4708 [nucl-th]} \BibitemShut
  {NoStop}%
\bibitem [{\citenamefont {Drescher}\ \emph {et~al.}(2006)\citenamefont
  {Drescher}, \citenamefont {Dumitru}, \citenamefont {Hayashigaki},\ and\
  \citenamefont {Nara}}]{Drescher:2006pi}%
  \BibitemOpen
  \bibfield  {author} {\bibinfo {author} {\bibfnamefont {H.-J.}\ \bibnamefont
  {Drescher}}, \bibinfo {author} {\bibfnamefont {A.}~\bibnamefont {Dumitru}},
  \bibinfo {author} {\bibfnamefont {A.}~\bibnamefont {Hayashigaki}}, \ and\
  \bibinfo {author} {\bibfnamefont {Y.}~\bibnamefont {Nara}},\ }\href@noop {}
  {\bibfield  {journal} {\bibinfo  {journal} {Phys. Rev.}\ }\textbf {\bibinfo
  {volume} {C74}} (\bibinfo {year} {2006})}\BibitemShut {NoStop}%
\bibitem [{\citenamefont {Adler}\ \emph {et~al.}(2014)\citenamefont {Adler}
  \emph {et~al.}}]{Adler:2013aqf}%
  \BibitemOpen
  \bibfield  {author} {\bibinfo {author} {\bibfnamefont {S.~S.}\ \bibnamefont
  {Adler}} \emph {et~al.} (\bibinfo {collaboration} {PHENIX}),\ }\href
  {\doibase 10.1103/PhysRevC.89.044905} {\bibfield  {journal} {\bibinfo
  {journal} {Phys. Rev.}\ }\textbf {\bibinfo {volume} {C89}},\ \bibinfo {pages}
  {044905} (\bibinfo {year} {2014})},\ \Eprint {http://arxiv.org/abs/1312.6676}
  {arXiv:1312.6676 [nucl-ex]} \BibitemShut {NoStop}%
\bibitem [{\citenamefont {Miller}\ \emph {et~al.}(2007)\citenamefont {Miller},
  \citenamefont {Reygers}, \citenamefont {Sanders},\ and\ \citenamefont
  {Steinberg}}]{Miller:2007ri}%
  \BibitemOpen
  \bibfield  {author} {\bibinfo {author} {\bibfnamefont {M.~L.}\ \bibnamefont
  {Miller}}, \bibinfo {author} {\bibfnamefont {K.}~\bibnamefont {Reygers}},
  \bibinfo {author} {\bibfnamefont {S.~J.}\ \bibnamefont {Sanders}}, \ and\
  \bibinfo {author} {\bibfnamefont {P.}~\bibnamefont {Steinberg}},\ }\href
  {\doibase 10.1146/annurev.nucl.57.090506.123020} {\bibfield  {journal}
  {\bibinfo  {journal} {Ann. Rev. Nucl. Part. Sci.}\ }\textbf {\bibinfo
  {volume} {57}},\ \bibinfo {pages} {205} (\bibinfo {year} {2007})},\ \Eprint
  {http://arxiv.org/abs/nucl-ex/0701025} {arXiv:nucl-ex/0701025 [nucl-ex]}
  \BibitemShut {NoStop}%
\bibitem [{\citenamefont {Shen}(2014)}]{Shen:2014sfi}%
  \BibitemOpen
  \bibfield  {author} {\bibinfo {author} {\bibfnamefont {C.}~\bibnamefont
  {Shen}},\ }\emph {\bibinfo {title} {{The standard model for relativistic
  heavy-ion collisions and electromagnetic tomography}}},\ \href@noop {} {Ph.D.
  thesis},\ \bibinfo  {school} {Ohio State University} (\bibinfo {year}
  {2014})\BibitemShut {NoStop}%
\bibitem [{\citenamefont {Adare}\ \emph {et~al.}(2008)\citenamefont {Adare}
  \emph {et~al.}}]{Adare:2008ns}%
  \BibitemOpen
  \bibfield  {author} {\bibinfo {author} {\bibfnamefont {A.}~\bibnamefont
  {Adare}} \emph {et~al.} (\bibinfo {collaboration} {PHENIX}),\ }\href
  {\doibase 10.1103/PhysRevC.78.044902} {\bibfield  {journal} {\bibinfo
  {journal} {Phys. Rev.}\ }\textbf {\bibinfo {volume} {C78}},\ \bibinfo {pages}
  {044902} (\bibinfo {year} {2008})},\ \Eprint {http://arxiv.org/abs/0805.1521}
  {arXiv:0805.1521 [nucl-ex]} \BibitemShut {NoStop}%
\bibitem [{\citenamefont {Dumitru}\ and\ \citenamefont
  {Nara}(2012)}]{Dumitru:2012yr}%
  \BibitemOpen
  \bibfield  {author} {\bibinfo {author} {\bibfnamefont {A.}~\bibnamefont
  {Dumitru}}\ and\ \bibinfo {author} {\bibfnamefont {Y.}~\bibnamefont {Nara}},\
  }\href {\doibase 10.1103/PhysRevC.85.034907} {\bibfield  {journal} {\bibinfo
  {journal} {Phys. Rev.}\ }\textbf {\bibinfo {volume} {C85}},\ \bibinfo {pages}
  {034907} (\bibinfo {year} {2012})},\ \Eprint {http://arxiv.org/abs/1201.6382}
  {arXiv:1201.6382 [nucl-th]} \BibitemShut {NoStop}%
\bibitem [{\citenamefont {Moreland}\ \emph {et~al.}(2013)\citenamefont
  {Moreland}, \citenamefont {Qiu},\ and\ \citenamefont
  {Heinz}}]{Moreland:2012qw}%
  \BibitemOpen
  \bibfield  {author} {\bibinfo {author} {\bibfnamefont {J.~S.}\ \bibnamefont
  {Moreland}}, \bibinfo {author} {\bibfnamefont {Z.}~\bibnamefont {Qiu}}, \
  and\ \bibinfo {author} {\bibfnamefont {U.~W.}\ \bibnamefont {Heinz}},\
  }\bibfield  {booktitle} {\emph {\bibinfo {booktitle} {{Proceedings, 23rd
  International Conference on Ultrarelativistic Nucleus-Nucleus Collisions :
  Quark Matter 2012 (QM 2012)}}},\ }\href {\doibase
  10.1016/j.nuclphysa.2013.02.141} {\bibfield  {journal} {\bibinfo  {journal}
  {Nucl. Phys.}\ }\textbf {\bibinfo {volume} {A904-905}},\ \bibinfo {pages}
  {815c} (\bibinfo {year} {2013})},\ \Eprint {http://arxiv.org/abs/1210.5508}
  {arXiv:1210.5508 [nucl-th]} \BibitemShut {NoStop}%
\bibitem [{\citenamefont {Bozek}\ and\ \citenamefont
  {Broniowski}(2013)}]{Bozek:2013uha}%
  \BibitemOpen
  \bibfield  {author} {\bibinfo {author} {\bibfnamefont {P.}~\bibnamefont
  {Bozek}}\ and\ \bibinfo {author} {\bibfnamefont {W.}~\bibnamefont
  {Broniowski}},\ }\href {\doibase 10.1103/PhysRevC.88.014903} {\bibfield
  {journal} {\bibinfo  {journal} {Phys. Rev.}\ }\textbf {\bibinfo {volume}
  {C88}},\ \bibinfo {pages} {014903} (\bibinfo {year} {2013})},\ \Eprint
  {http://arxiv.org/abs/1304.3044} {arXiv:1304.3044 [nucl-th]} \BibitemShut
  {NoStop}%
\bibitem [{\citenamefont {Ansorge}\ \emph {et~al.}(1989)\citenamefont {Ansorge}
  \emph {et~al.}}]{Ansorge:1988kn}%
  \BibitemOpen
  \bibfield  {author} {\bibinfo {author} {\bibfnamefont {R.~E.}\ \bibnamefont
  {Ansorge}} \emph {et~al.} (\bibinfo {collaboration} {UA5}),\ }\bibfield
  {booktitle} {\emph {\bibinfo {booktitle} {{In *Munich 1988, Proceedings, High
  energy physics* 647-648.}}},\ }\href {\doibase 10.1007/BF01506531} {\bibfield
   {journal} {\bibinfo  {journal} {Z. Phys.}\ }\textbf {\bibinfo {volume}
  {C43}},\ \bibinfo {pages} {357} (\bibinfo {year} {1989})}\BibitemShut
  {NoStop}%
\bibitem [{\citenamefont {Adler}\ \emph {et~al.}(2004)\citenamefont {Adler}
  \emph {et~al.}}]{Adler:2003cb}%
  \BibitemOpen
  \bibfield  {author} {\bibinfo {author} {\bibfnamefont {S.~S.}\ \bibnamefont
  {Adler}} \emph {et~al.} (\bibinfo {collaboration} {PHENIX}),\ }\href@noop {}
  {\bibfield  {journal} {\bibinfo  {journal} {Phys. Rev.}\ }\textbf {\bibinfo
  {volume} {C69}},\ \bibinfo {pages} {034909} (\bibinfo {year} {2004})},\
  \Eprint {http://arxiv.org/abs/nucl-ex/0307022} {nucl-ex/0307022} \BibitemShut
  {NoStop}%
\bibitem [{\citenamefont {Adare}\ \emph
  {et~al.}(2015{\natexlab{b}})\citenamefont {Adare} \emph
  {et~al.}}]{Adare:2015bcj}%
  \BibitemOpen
  \bibfield  {author} {\bibinfo {author} {\bibfnamefont {A.}~\bibnamefont
  {Adare}} \emph {et~al.} (\bibinfo {collaboration} {PHENIX}),\ }\href
  {\doibase 10.1103/PhysRevC.92.034914} {\bibfield  {journal} {\bibinfo
  {journal} {Phys. Rev.}\ }\textbf {\bibinfo {volume} {C92}},\ \bibinfo {pages}
  {034914} (\bibinfo {year} {2015}{\natexlab{b}})},\ \Eprint
  {http://arxiv.org/abs/1504.05168} {arXiv:1504.05168 [nucl-ex]} \BibitemShut
  {NoStop}%
\bibitem [{\citenamefont {Adamczyk}\ \emph {et~al.}(2013)\citenamefont
  {Adamczyk} \emph {et~al.}}]{Adamczyk:2013jh}%
  \BibitemOpen
  \bibfield  {author} {\bibinfo {author} {\bibfnamefont {L.}~\bibnamefont
  {Adamczyk}} \emph {et~al.} (\bibinfo {collaboration} {STAR}),\ }\href@noop {}
  {\bibfield  {journal} {\bibinfo  {journal} {Phys. Rev. C}\ }\textbf {\bibinfo
  {volume} {88}},\ \bibinfo {pages} {014904} (\bibinfo {year}
  {2013})}\BibitemShut {NoStop}%
\bibitem [{\citenamefont {Adare}\ \emph {et~al.}(2011)\citenamefont {Adare}
  \emph {et~al.}}]{Adare:2011uo}%
  \BibitemOpen
  \bibfield  {author} {\bibinfo {author} {\bibfnamefont {A.}~\bibnamefont
  {Adare}} \emph {et~al.} (\bibinfo {collaboration} {PHENIX}),\ }\href@noop {}
  {\bibfield  {journal} {\bibinfo  {journal} {Phys. Rev. Lett.}\ }\textbf
  {\bibinfo {volume} {107}},\ \bibinfo {pages} {252301} (\bibinfo {year}
  {2011})}\BibitemShut {NoStop}%
\end{thebibliography}%

\end{document}